\documentclass[english,11pt,a4paper]{article}
\usepackage[T1]{fontenc}
\usepackage[latin9]{inputenc}
\usepackage{booktabs}
\usepackage{amsmath}
\usepackage{amsthm}
\usepackage{amssymb}
\usepackage{graphicx}
\usepackage{wasysym}

\makeatletter

\providecommand{\tabularnewline}{\\}

\numberwithin{equation}{section}

\usepackage{jheppub}
\usepackage{slashed}

\usepackage{tikz}
\usetikzlibrary{shapes,arrows}
\usetikzlibrary{decorations.pathmorphing,decorations.markings}
\usetikzlibrary{decorations.pathreplacing,calligraphy}
\usetikzlibrary{snakes}

\allowdisplaybreaks

\makeatother

\usepackage{babel}
\begin{document}
\title{ Production rates of dark photons and $Z'$ in the Sun and
 stellar cooling bounds}

\abstract{ Light weakly interacting particles could be copiously
produced in the Sun which, as a well-understood star, could provide
severe constraints on such new physics. In this work, we calculate
the solar production rates of light gauge bosons (e.g.~dark photon)
arising from various $U(1)$ extensions of the standard model. It
is known that the dark photon production rate is suppressed by the
dark photon mass if it is well below the plasmon mass of the medium.
We show that for more general $U(1)$ gauge bosons, this suppression
is absent if the couplings are not in alignment with those of the
photon. We investigate a few frequently discussed $U(1)$ models including
 $B-L$, $L_{\mu}-L_{\tau}$, and $L_{e}-L_{\mu(\tau)}$, and derive
the stellar cooling bounds for these models. }

\author[a]{Shao-Ping Li} 
\author[a]{and Xun-Jie Xu} 
\affiliation[a]{Institute of High Energy Physics, Chinese Academy of Sciences, Beijing 100049, China} 
\preprint{\today}  
\emailAdd{spli@ihep.ac.cn} 
\emailAdd{xuxj@ihep.ac.cn}

\preprint{\today}
\maketitle

\section{Introduction}

New physics beyond the Standard Model (SM) may be hidden at low-energy
scales with weak couplings to the SM.  Dark photons or, more generally,
 dark neutral gauge bosons arising from $U(1)$ extensions of the
SM are extensively studied cases of such new physics,   with rising
interest recently at the intensity frontier\footnote{See~\cite{Jaeckel:2010ni,Essig:2013lka,Alexander:2016aln,Ilten:2018crw,Bauer:2018onh,Fabbrichesi:2020wbt}
for reviews. In particular, searching for dark photons has been one
of the major scientific goals of many ongoing or upcoming experiments
like Belle-II~\cite{Belle-II:2022jyy}, FASER~\cite{Feng:2017uoz,FASER:2023zcr},
SHiP~\cite{Alekhin:2015byh}, SeaQuest~\cite{Gardner:2015wea,Berlin:2018pwi},
MATHUSLA~\cite{Chou:2016lxi,MATHUSLA:2020uve}, etc. } and in cosmology~\cite{Fradette:2014sza,Berger:2016vxi,Knapen:2017xzo,Agrawal:2018vin,McDermott:2019lch,Coffey:2020oir,Ferreira:2020fam}.
  Notably, they are also considered as popular dark matter candidates
with interesting low-energy observational consequences~\cite{An:2020jmf,An:2023wij}.

Hot and dense astrophysical environments can be used to probe light
and weakly-coupled particles if they are copiously produced via frequent
collisions of medium particles.  In well-modeled stars such as the
Sun, the additional energy loss caused by the emission of such new
particles has been used to impose one of the  most restrictive bounds
on  them in certain mass ranges~\cite{Raffelt:1987yu,Raffelt:1990yz,Raffelt1996,Raffelt:1999tx,Davidson:2000hf,Redondo:2008aa,An:2013yfc,Redondo:2013lna,Vinyoles:2015aba,Hardy:2016kme,Chu:2019rok,DeRocco:2020xdt,An:2020bxd,Carenza:2020zil,Capozzi:2020cbu,Balaji:2022noj,Yamamoto:2023zlu,Bottaro:2023gep}.
  In supernovae and neutron stars, new particles of higher masses
could be probed~\cite{Raffelt1996,Raffelt:1999tx,Davidson:2000hf,Dent:2012mx,Kazanas:2014mca,Rrapaj:2015wgs,Chang:2016ntp,Heurtier:2016otg,Dev:2020eam,Hong:2020bxo,Shin:2021bvz,Shin:2022ulh},
though the astrophysical models have more diverse and larger uncertainties. 

It is known that the production rate of a very light dark photon in
a thermal medium is suppressed by its mass~\cite{Redondo:2008aa,An:2013yfc,Redondo:2013lna}.
In fact, in the massless limit due to  mass degeneracy with the SM
photon, one can always find a basis in which the dark photon is fully
decoupled from the SM.  Hence, the low-mass suppression is expected. 

In this work, we would like to point out that the low-mass suppression
is a generic feature for any gauge bosons with photon-like couplings,
i.e., couplings to medium particles being proportional to their electric
charges.  However, if the couplings are not photon-like,   which
is also common in many $U(1)$ extensions of the SM (e.g.~$B-L$~\cite{Davidson:1978pm,Mohapatra:1980qe,Wetterich:1981bx},
$L_{e}-L_{\mu}$~\cite{Foot:1990mn,He:1990pn,He:1991qd}), then the
low-mass suppression is absent. In this case, the production rate
becomes independent of the mass in the low-mass limit. It is worth
mentioning that even for the dark photon model the couplings should
slightly deviate from the aforementioned photon-like scenario\footnote{When considering the SM gauge invariance which only allows for kinetic
mixing with the SM $U(1)_{Y}$ gauge field, the dark photon should
also slightly mix with the $Z$ boson, which causes the small deviation---see
Sec.~\ref{subsec:Two-limits} for a detailed discussion. }, though the effect of this deviation turns out to be negligible.

In addition to the dark photon model, we investigate a few popular
$U(1)$ extensions including $B-L$, $L_{\mu}-L_{\tau}$, $L_{e}-L_{\mu}$,
and $L_{e}-L_{\tau}$ which  have been frequently considered in the
literature~ in recent years~\cite{Heeck:2011wj,Harnik:2012ni,Heeck:2014zfa,Crivellin:2015mga,Altmannshofer:2016jzy,Wise:2018rnb,Kamada:2018zxi,Asai:2018ocx,Lindner:2018kjo,Heeck:2018nzc,Escudero:2019gzq,Asai:2019ciz,Esmaili:2019pcy,Chen:2020jvl,Huang:2021nkl}.
  Among these $U(1)$ models, the low-mass suppression is present
for $L_{\mu}-L_{\tau}$ but not for $B-L$ and $L_{e}-L_{\mu(\tau)}$.
Consequently, the stellar cooling bounds for these models have their
respective very different low-mass limits.  

This paper is organized as follows. In Sec.~\ref{sec:model}, we
formulate the most general $U(1)$ extension of the SM, derive interactions
of the new gauge boson (generically denoted by $Z'$) in the physical
basis, and discuss when the $Z'$ features photon-like couplings.
  In Sec.~\ref{sec:Production}, we calculate the production rate
of $Z'$ in the solar medium, with the focus on when and how the rate
would vanish in the massless limit. Then in Sec.~\ref{sec:bounds},
we apply the calculations to a few $U(1)$ models and derive the corresponding
stellar coupling bounds. Finally we conclude in Sec.~\ref{sec:Con}
and relegate some details to the appendix.

\section{Generic dark gauge boson\label{sec:model}}

In our framework, we consider a generic dark gauge boson arising from
a generic $U(1)$ extension of the SM, $SU(3)_{c}\times SU(2)_{L}\times U(1)_{Y}\times U(1)_{X}$.
Such a gauge boson is called the dark photon if it couples to the
SM content only via the kinetic mixing with the photon~\cite{Holdom:1985ag}.
It is also generically referred to as $Z'$, defined rather broadly
as a new neutral gauge boson similar to the SM $Z$ boson. Although
for historical reasons $Z'$ is often considered to be around or above
the electroweak scale~\cite{Langacker:2008yv}, it can be light
as well.  We regard the dark photon as a special case of $Z'$ and
throughout this paper, use $Z'$ as a  more generic name and notation
inclusively.

\subsection{Interactions in the original and physical bases}

In the original basis, the Lagrangian of relevant kinetic terms reads:
\begin{eqnarray}
{\cal L} & \supset & -\frac{1}{4}W_{\mu\nu}W^{\mu\nu}-\frac{1}{4}\hat{B}_{\mu\nu}\hat{B}^{\mu\nu}-\frac{1}{4}\hat{X}_{\mu\nu}\hat{X}^{\mu\nu}-\frac{\epsilon}{2}\hat{B}_{\mu\nu}\hat{X}^{\mu\nu}+\sum_{\psi}\overline{\psi}\gamma^{\mu}iD_{\mu}\psi\thinspace,\label{eq:L}
\end{eqnarray}
where $W$, $B$, and $X$ denote the gauge field strength tensors
of $SU(2)_{L}$, $U(1)_{Y}$, and $U(1)_{X}$, respectively. We add
hats ``$\hat{\ \ }$'' on two of them to remind the reader that
their kinetic terms are not canonical in the original  basis. The
last term of Eq.~\eqref{eq:L} in which $\psi$ is a SM fermion gives
rise to the interactions of gauge bosons with fermions. The covariant
derivative is defined as 
\begin{equation}
D_{\mu}\equiv\partial_{\mu}-ig\sum_{a=1}^{3}\frac{\sigma_{a}}{2}W_{a,\mu}-ig'Q_{Y}\hat{B}_{\mu}-ig_{Z'}Q_{X}\hat{X}_{\mu}\thinspace,\label{eq:D}
\end{equation}
where $Q_{Y}$ and $Q_{X}$ denote the charges of $\psi$ under $U(1)_{Y}$
and $U(1)_{X}$. 

Since the kinetic terms are not canonical due to the $\epsilon\hat{B}_{\mu\nu}\hat{X}^{\mu\nu}$
term, one needs to perform a linear (non-unitary) transformation to
canonicalize the kinetic terms, for the applicability of the standard
Feynman rules. We denote the fields after such a transformation by
notations without hats ($\hat{X}\to X$, $\hat{B}\to B$). On the
other hand, similar to the well-known fact of the SM that  $W_{3}$
and $B$ are not in the mass eigenbasis, here mass mixing among $W$,
$B$, and $X$ after symmetry breaking is generally expected. So a
subsequent unitary (orthogonal) transformation is needed to obtain
the physical states of gauge bosons with well-defined masses. Among
them, there must be a massless state, which  is defined as the photon,
$A$.\footnote{Since it is also conventional to denote the photon by $\gamma$, we
use both $\gamma$ and $A$ interchangeably. The latter is more appropriate
if Lorentz indices need to be explicit (e.g.~$A^{\mu}$) and the
former is more commonly used in reaction processes (e.g.~$e^{-}+p\to e^{-}+p+\gamma$). } There should also be two massive neutral states. We denote them by
$Z$ and $Z'$, their masses by $m_{Z}$ and $m_{Z'}$, and the ratio
$r_{m}\equiv m_{Z'}^{2}/m_{Z}^{2}$. 

We refer to the basis in which all gauge bosons are mass eigenstates
with canonical kinetic terms as\emph{ the physical basis.}  It is
connected to the original basis via 
\begin{equation}
(W_{1},\ W_{2},\ A,\ Z,\ Z')^{T}=\mathbb{T}(W_{1},\ W_{2},\ W_{3},\ \hat{B},\ \hat{X})^{T},\label{eq:-30}
\end{equation}
where $\mathbb{T}$ is a  $5\times5$ matrix combining the aforementioned
transformations. Its specific form is derived and presented in Appendix~\ref{sec:Transformation}. 

\begin{table*}
% \resizebox{\textwidth}{!}{%Scale down your table to the textwidth
\newcommand{\vvvv}{ \rule[-2ex]{0pt}{5ex} }\centering

\begin{tabular}{cccc}
\toprule 
$\psi$ & $g_{Z}^{(\psi)}$ & $g_{Z'}^{(\psi)}$ & \tabularnewline
\midrule 
\vvvv $e_{L}$ & $\frac{g}{c_{W}}\left(s_{W}^{2}-\frac{1}{2}\right)+\epsilon g_{X}^{(\psi)}\frac{s_{W}}{1-r_{m}}$ & $g_{X}^{(\psi)}+\epsilon gs_{W}\frac{2c_{W}^{2}-r_{m}}{2c_{W}\left(1-r_{m}\right)}$ & \tabularnewline
\vvvv $e_{R}$ & $\frac{g}{c_{W}}\left(s_{W}^{2}-\frac{1}{2}\right)+\epsilon g_{X}^{(\psi)}\frac{s_{W}}{1-r_{m}}$ & $g_{X}^{(\psi)}+\epsilon gs_{W}\frac{c_{W}^{2}-r_{m}}{c_{W}\left(1-r_{m}\right)}$ & \tabularnewline
\vvvv $u_{L}$ & $\frac{2g}{3c_{W}}\left(\frac{3}{4}-s_{W}^{2}\right)+\epsilon g_{X}^{(\psi)}\frac{s_{W}}{1-r_{m}}$ & $g_{X}^{(\psi)}-\epsilon gs_{W}\frac{4c_{W}^{2}-r_{m}}{6c_{W}\left(1-r_{m}\right)}$ & \tabularnewline
\vvvv $u_{R}$ & $\frac{2g}{3c_{W}}\left(-s_{W}^{2}\right)+\epsilon g_{X}^{(\psi)}\frac{s_{W}}{1-r_{m}}$ & $g_{X}^{(\psi)}-\epsilon gs_{W}\frac{2\left(c_{W}^{2}-r_{m}\right)}{3c_{W}\left(1-r_{m}\right)}$ & \tabularnewline
\vvvv $d_{L}$ & $\frac{g}{3c_{W}}\left(s_{W}^{2}-\frac{3}{2}\right)+\epsilon g_{X}^{(\psi)}\frac{s_{W}}{1-r_{m}}$ & $g_{X}^{(\psi)}+\epsilon gs_{W}\frac{2c_{W}^{2}+r_{m}}{6c_{W}\left(1-r_{m}\right)}$ & \tabularnewline
\vvvv $d_{R}$ & $\frac{g}{3c_{W}}s_{W}^{2}+\epsilon g_{X}^{(\psi)}\frac{s_{W}}{1-r_{m}}$ & $g_{X}^{(\psi)}+\epsilon gs_{W}\frac{c_{W}^{2}-r_{m}}{3c_{W}\left(1-r_{m}\right)}$ & \tabularnewline
\vvvv $\nu_{L}$ & $\frac{g}{2c_{W}}+\epsilon g_{X}^{(\psi)}\frac{s_{W}}{1-r_{m}}$ & $g_{X}^{(\psi)}-\epsilon gs_{W}\frac{r_{m}}{2c_{W}\left(1-r_{m}\right)}$ & \tabularnewline
\vvvv $\nu_{R}$ & $\epsilon g_{X}^{(\psi)}\frac{s_{W}}{1-r_{m}}$ & $g_{X}^{(\psi)}$ & \tabularnewline
\bottomrule
\end{tabular}

% }

\caption{\label{tab:g-eff}Effective couplings of $Z$ and $Z'$ in the $SU(2)_{L}\times U(1)_{Y}\times U(1)_{X}$
model assuming that the kinetic mixing term is the only source of
mass mixing. Here $r_{m}\equiv m_{Z'}^{2}/m_{Z}^{2}$, $g_{X}^{(\psi)}\equiv g_{Z'}Q_{X}^{(\psi)}$
with $g_{Z'}$ and $Q_{X}^{(\psi)}$ defined in Eq.~\eqref{eq:D}.
Since some models such as $U(1)_{B-L}$ may introduce right-handed
neutrinos ($\nu_{R}$), we also include $\nu_{R}$ here for completeness. }
\end{table*}

Applying the transformation \eqref{eq:-30} to Eq.~\eqref{eq:D},
one obtains the gauge interactions in the physical basis. According
to Appendix~\ref{sec:Transformation}, the gauge interactions of
$A$ (photon) remain exactly the same as the SM ones, while for $Z$
and $Z'$, the gauge interactions are given by
\begin{equation}
{\cal L}\supset g_{Z}^{(\psi)}\overline{\psi}\slashed{Z}\psi+g_{Z'}^{(\psi)}\overline{\psi}\slashed{Z}'\psi\thinspace,\label{eq:-31}
\end{equation}
where $g_{Z}^{(\psi)}$ and $g_{Z'}^{(\psi)}$ are listed in Tab.~\ref{tab:g-eff},
assuming the kinetic mixing term is the only source of the mass mixing.
One should note that this assumption is valid only when $Z'$ obtains
its mass via the Stückelberg mechanism~\cite{Stueckelberg:1938hvi,Feldman:2007wj}
or the Higgs mechanism with SM singlet scalars that are only charged
under $U(1)_{X}$. For the Higgs mechanism involving e.g.~Higgs doublets
charged under both $U(1)_{X}$ and $SU(2)_{L}\times U(1)_{Y}$, $g_{Z}^{(\psi)}$
and $g_{Z'}^{(\psi)}$ would involve an additional independent mass
mixing parameter $\theta$. We refer to Eqs.~\eqref{eq:-9} and \eqref{eq:-10}
in Appendix~\ref{sec:Transformation} for the results in this case.

\subsection{The dark photon model: the photon-like and the  hypercharge limits
\label{subsec:Two-limits}}

In our work, the dark photon model is defined as a special case of
the above model: either $Q_{X}^{(\psi)}=0$ or $g_{Z'}$ is negligibly
small. Hence the gauge interactions in the physical basis can be obtained
by taking $g_{X}^{(\psi)}\equiv g_{Z'}Q_{X}^{(\psi)}\to0$ in Tab.~\ref{tab:g-eff}.
   Here we would like to outline two interesting limits in the
dark photon model.
\begin{itemize}
\item The photon-like limit. For a very light $Z'$, we take the limit $r_{m}\to0$
in Tab.~\ref{tab:g-eff} with $g_{X}^{(\psi)}=0$ and obtain
\begin{equation}
g_{Z'}^{(\psi)}=-\epsilon ec_{W}Q_{{\rm em}}^{(\psi)}\thinspace,\ \ \text{i.e.}\ g_{Z'}^{(\psi)}\propto Q_{{\rm em}}^{(\psi)}\thinspace,\label{eq:-11}
\end{equation}
where $e=gs_{W}$  and $Q_{{\rm em}}^{(\psi)}$ is the electric charge
of $\psi$.  Eq.~\eqref{eq:-11} implies that in the zero mass limit,
the interactions of $Z'$ with fermions are very similar to the interactions
of the photon in the sense that $\left(g_{Z'}^{(e)}:g_{Z'}^{(\nu)}:g_{Z'}^{(u)}:g_{Z'}^{(d)}\right)=\left(-1:0:2/3:-1/3\right)$.
\item The hypercharge limit. For a heavy $Z'$ with the mass well above
the electroweak scale, we take the limit $r_{m}\to\infty$ together
with $g_{X}^{(\psi)}=0$ in Tab.~\ref{tab:g-eff} and obtain
\begin{equation}
g_{Z'}^{(\psi)}=\epsilon g\frac{s_{W}}{c_{W}}Q_{Y}^{(\psi)}\thinspace,\ \ \text{i.e.}\ g_{Z'}^{(\psi)}\propto Q_{Y}^{(\psi)}\thinspace.\label{eq:-11-2}
\end{equation}
For example, in this limit we have $\left(g_{Z'}^{(e_{L},\nu_{L})}:g_{Z'}^{(e_{R})}:g_{Z'}^{(u_{L},d_{L})}\right)=\left(-1/2:-1:1/6\right)$.
This is expected because at high energies the electroweak symmetry
is restored and $Z'$ essentially mixes only with the SM $U(1)_{Y}$
gauge field. 
\end{itemize}

\subsection{The simplified dark photon model: the decoupling limit\label{subsec:decouple}}

In the literature, it is quite common that only the kinetic mixing
between $\hat{X}$ and the photon ($\hat{A}$) is considered:
\begin{equation}
{\cal L}\supset-\frac{1}{4}\hat{A}_{\mu\nu}\hat{A}^{\mu\nu}-\frac{1}{4}\hat{X}_{\mu\nu}\hat{X}^{\mu\nu}-\frac{\varepsilon}{2}\hat{A}_{\mu\nu}\hat{X}^{\mu\nu}+eJ_{{\rm em}}^{\mu}\hat{A}_{\mu}+\frac{1}{2}m_{X}^{2}\hat{X}_{\mu}\hat{X}^{\mu}\thinspace,\label{eq:simp}
\end{equation}
where $\varepsilon\equiv\epsilon c_{W}$ and $J_{{\rm em}}^{\mu}$
is the electromagnetic current. This Lagrangian is not gauge invariant
and we refer to it as \emph{the simplified dark photon model}. Disregarding
the gauge invariance issue, it can concisely capture the main feature
of the complete model in the low-$m_{Z'}$ limit, i.e.~the photon-like
limit in Sec.~\ref{subsec:Two-limits}. 

In the simplified dark photon model, the transformation in Eq.~\eqref{eq:-30}
is reduced to $(A,\ Z')^{T}=\mathbb{T}(\hat{A},\ \hat{X})^{T}$ with
\begin{equation}
\mathbb{T}=\left(\begin{array}{cc}
1 & \varepsilon\\
0 & \sqrt{1-\varepsilon^{2}}
\end{array}\right).\label{eq:-32}
\end{equation}
And the kinetic, mass, and gauge interaction terms are transformed
as follows:
\begin{align}
\text{kinetic matrix}:\  & K=\left(\begin{array}{cc}
1 & \varepsilon\\
\varepsilon & 1
\end{array}\right)\xrightarrow{\mathbb{T}}\left(\begin{array}{cc}
1 & 0\\
0 & 1
\end{array}\right),\label{eq:-33}\\
\text{mass matrix}:\  & M^{2}=\left(\begin{array}{cc}
0 & 0\\
0 & m_{X}^{2}
\end{array}\right)\xrightarrow{\mathbb{T}}\left(\begin{array}{cc}
0 & 0\\
0 & m_{Z'}^{2}
\end{array}\right),\ m_{Z'}^{2}=\frac{m_{X}^{2}}{1-\varepsilon^{2}}\thinspace,\label{eq:-34}\\
\text{current matrix}:\  & J=\left(\begin{array}{c}
J_{{\rm em}}\\
0
\end{array}\right)\xrightarrow{\mathbb{T}}J_{{\rm em}}\left(\begin{array}{c}
1\\
x
\end{array}\right),\ x=\frac{-\varepsilon}{\sqrt{1-\varepsilon^{2}}}\thinspace,\label{eq:-35}
\end{align}
where matrices $K$, $M^{2}$ and $J$ are defined by rewriting Eq.~\eqref{eq:simp}
as ${\cal L}\supset-\frac{1}{4}V_{\mu\nu}^{T}KV^{\mu\nu}+J_{\mu}^{T}V^{\mu}+\frac{1}{2}V_{\mu}^{T}M^{2}V^{\mu}$
with $V_{\mu}=(\hat{A}_{\mu},\ \hat{X}_{\mu})^{T}$. 

It is important to notice that $\mathbb{T}$ is unique if one requires
that it simultaneously diagonalizes $K$ and $M^{2}$. However, if
$m_{X}^{2}=0$, then $\mathbb{T}$ is no longer unique, because if
$\mathbb{T}\to\mathbb{T}'=\mathbb{O}\cdot\mathbb{T}$ with $\mathbb{O}$
an arbitrary orthogonal matrix ($\mathbb{O}\cdot\mathbb{O}^{T}=1$),
Eq.~\eqref{eq:-33} remains the same while $M^{2}=0$ is unchanged.
In other words, the physical basis in which all gauge bosons are mass
eigenstates is no longer unique due to the mass degeneracy. In this
particular case, one could further choose an appropriate $\mathbb{O}$
so that
\begin{equation}
K=\mathbb{T}'^{T}\left(\begin{array}{cc}
1 & 0\\
0 & 1
\end{array}\right)\mathbb{T}',\ M^{2}=0,\ J\propto\mathbb{T}'^{T}\left(\begin{array}{c}
1\\
0
\end{array}\right),\label{eq:-36}
\end{equation}
which implies that under this basis, the dark degree of freedom is
fully decoupled from the SM. We refer to this as the decoupling limit.
Since  physical results should be basis independent, when computing
e.g.~stellar energy loss rates in other bases where $Z'$ is superficially
coupled to $J_{{\rm em}}$, some cancellations in the final results
are expected, as will be shown explicitly in our calculation in Sec.~\ref{subsec:vanishing}. 

A further generalization, which to our knowledge has not been noticed
in the literature, is that even if the lower component of $J$ in
Eq.~\eqref{eq:-35} is nonzero, one can still rotate it away in the
zero mass limit so that a massless $Z'$ is fully decoupled. This
is possible if $\hat{X}$ is originally coupled to SM fermions ($Q_{X}^{(\psi)}\neq0$)
but the corresponding current, $J_{X}$, is proportional to $J_{{\rm em}}$
(i.e.~$J_{X}^{\mu}\propto2/3\overline{u}\gamma^{\mu}u-1/3\overline{d}\gamma^{\mu}d-\overline{e}\gamma^{\mu}e+\cdots$).
This will also be shown in the next section.

\section{\label{sec:Production}Production of $Z'$ in the Sun }

\subsection{The vanishing production rate of a photon-like\label{subsec:vanishing}
$Z'$}

It is known that the dark photon production rate in a finite-temperature
environment vanishes in the limit of zero mass~\cite{Redondo:2008aa,An:2013yfc,Redondo:2013lna,Hardy:2016kme}.
This feature can be understood from the decoupling limit discussed
in Sec.~\ref{subsec:decouple}. 

Here we would like to make a generalization that this is also true
for other $Z'$ with photon-like couplings. For instance, in plasma
consisting of only protons and electrons, the $Z'$ in the $U(1)_{B-L}$
model has $g_{Z'}^{(p)}=-g_{Z'}^{(e)}$, which makes the relevant
couplings photon-like.  Consequently, its production rate vanishes
in the $m_{Z'}\to0$ limit~\cite{Hardy:2016kme}.  

Below we show explicitly how it vanishes for a photon-like $Z'$.
In previous studies, this feature is shown for the dark photon in
the basis of $\hat{\mathbf{V}}$ where the  gauge bosons are not in
mass eigenstates. Here we adopt the physical basis ($\mathbf{V}_{{\rm ph}}$)
for a rigorous treatment. 

\begin{figure}
\centering

\includegraphics[width=0.6\textwidth]{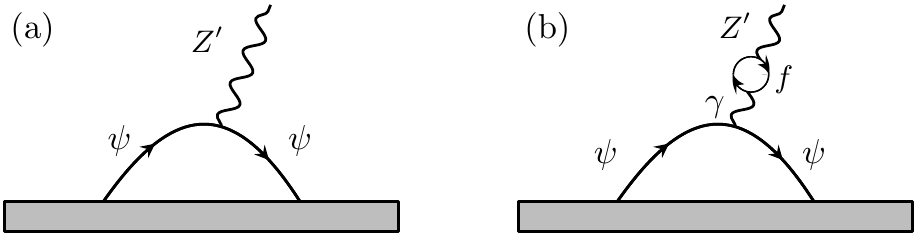}

\caption{\label{fig:cancel}A generic diagram for $Z'$ production (left) accompanied
with another diagram (right) which is equally important in a thermal
or dense medium. The gray boxes represent remaining parts of the diagrams,
which can be quite arbitrary. }
\end{figure}

Consider a generic Feynman diagram for $Z'$ production, as shown
in the left panel of Fig.~\ref{fig:cancel}. Since in the physical
basis  there is no mixing between any gauge bosons, the final-state
$Z'$ has to be attached to a fermion $\psi$ in the diagram. Hence
the diagram should be proportional to the coupling $g_{Z'}^{(\psi)}$. 

On the other hand, such a diagram is always accompanied with another
diagram shown in the right panel of Fig.~\ref{fig:cancel}. This
diagram would be just a higher-order correction in vacuum, but in
a thermal or dense medium it can be equally important. It is known
that coherent scattering of photons with charged particles in a medium
modifies the photon dispersion relation. In plasma, this corresponds
to the well-known plasmon masses. In a transparent medium, this gives
rise to the refractive index. The medium  effect can be calculated
by evaluating the self-energy loop of the photon using finite-temperature/density
field theories. Alternatively, one can compute the medium effect
using the coherent scattering theory---see Appendix~\ref{sec:coherent-scattering}
for a re-derivation. If the final-state photon is changed to $Z'$,
one obtains medium-induced mixing between $Z'$ and the photon~\cite{Raffelt:1987im}.
\emph{Therefore, even though we have removed all mixing between gauge
bosons in vacuum, the medium effect still causes additional mixing}. 

The medium-induced mixing between $\gamma$ and $Z'$  can be computed
by rescaling the self-energy of the photon in the medium:
\begin{equation}
\Pi_{\gamma-(f)-Z'}^{\mu\nu}=\frac{g_{Z'}^{(f)}}{eQ_{{\rm em}}^{(f)}}\Pi_{\gamma-(f)-\gamma}^{\mu\nu}\thinspace,\label{eq:-14}
\end{equation}
where $\Pi_{\gamma-(f)-\gamma}^{\mu\nu}$ denotes the photon self-energy,
with $f$ the charged fermion running in the loop. The self-energy
$\Pi_{\gamma-(f)-\gamma}^{\mu\nu}$ is conventionally decomposed as
 
\begin{equation}
\Pi_{\gamma-(f)-\gamma}^{\mu\nu}=\Pi_{\gamma-(f)-\gamma}^{L}\epsilon_{L}^{\mu}\epsilon_{L}^{*\nu}+\Pi_{\gamma-(f)-\gamma}^{T}\left(\epsilon_{T1}^{\mu}\epsilon_{T1}^{*\nu}+\epsilon_{T2}^{\mu}\epsilon_{T2}^{*\nu}\right),\label{eq:-16}
\end{equation}
where $\epsilon_{L}$ and $\epsilon_{T1,2}$ are the longitudinal
and two transverse polarization vectors, and $\Pi_{\gamma-(f)-\gamma}^{L,T}$
in non-degenerate and non-relativistic plasma can be found in Refs.~\cite{Braaten:1993jw,Raffelt1996,Hardy:2016kme}\footnote{Note that at the one-loop level, $\Pi_{\gamma-(f)-\gamma}^{L,T}$
are real. Their imaginary parts, which are related to the thermal
production rate of $\gamma$, only arise at two-loop or higher levels. }:
\begin{equation}
\Pi_{\gamma-(f)-\gamma}^{T}\approx Q_{f}^{2}\frac{e^{2}n_{f}}{m_{f}}\thinspace,\ \ \Pi_{\gamma-(f)-\gamma}^{L}\approx Q_{f}^{2}\left(1-\frac{|\mathbf{k}|^{2}}{\omega^{2}}\right)\frac{e^{2}n_{f}}{m_{f}}\thinspace.\label{eq:-15}
\end{equation}
Here the photon momentum is $k^{\mu}=(\omega,\ \mathbf{k})$; $n_{f}$,
$m_{f}$, and $Q_{f}$ are the number density, mass, and electric
charge of $f$,\footnote{Here the fermion $f$ should be unbound particles (such as electrons
and ions in the plasma). Charged particles in bound states such as
quarks in protons or protons in helium should not be taken into account
individually. } respectively. 

Now applying the in-medium photon self-energy and $\gamma$-$Z'$
mixing to the second diagram in Fig.~\ref{fig:cancel}, we obtain
\begin{equation}
\text{diagram (b)}\propto eQ_{{\rm em}}^{(\psi)}\cdot\frac{-i}{k^{2}-\Pi_{\gamma\gamma}^{L,T}}\cdot i\Pi_{\gamma Z'}^{L,T}\thinspace,\label{eq:-17}
\end{equation}
where 
\begin{equation}
\Pi_{\gamma\gamma}^{L,T}\equiv\sum_{f}\Pi_{\gamma-(f)-\gamma}^{L,T},\ \ \Pi_{\gamma Z'}^{L,T}\equiv\sum_{f}\Pi_{\gamma-(f)-Z'}^{L,T}\thinspace.\label{eq:-18}
\end{equation}

For photon-like $Z'$ couplings, $g_{Z'}^{(f)}\propto eQ_{{\rm em}}^{(f)}$,
let us  define a universal ratio $r_{g}$ between them, 
\begin{equation}
g_{Z'}^{(f)}=r_{g}\thinspace eQ_{{\rm em}}^{(f)}\thinspace,\label{eq:-19}
\end{equation}
so that $\Pi_{\gamma Z'}^{L,T}=r_{g}\Pi_{\gamma\gamma}^{L,T}$. Therefore,
the two diagrams in Fig.~\ref{fig:cancel} can be combined to give
\begin{align}
\text{diagrams (a) + (b)} & \propto g_{Z'}^{(\psi)}+eQ_{{\rm em}}^{(\psi)}\frac{r_{g}\Pi_{\gamma\gamma}^{L,T}}{k^{2}-\Pi_{\gamma\gamma}^{L,T}}\nonumber \\
 & \propto g_{Z'}^{(\psi)}\left[1+\frac{\Pi_{\gamma\gamma}^{L,T}}{m_{Z'}^{2}-\Pi_{\gamma\gamma}^{L,T}}\right]\nonumber \\
 & \propto g_{Z'}^{(\psi)}\frac{m_{Z'}^{2}}{m_{Z'}^{2}-\Pi_{\gamma\gamma}^{L,T}}\thinspace,\label{eq:-20}
\end{align}
which implies that the two diagrams cancel out in the $m_{Z'}^{2}\to0$
limit, provided that $\Pi_{\gamma\gamma}^{L,T}$ is finite. This conclusion
only relies on Eq.~\eqref{eq:-19}, irrespective of the specific forms
of $\Pi_{\gamma\gamma}^{L,T}$.

\subsection{The production rate of a generic $Z'$}

For a generic $Z'$ with the couplings disproportional to the electric
charges, the aforementioned cancellation is generally absent. For
simplicity, let us first consider ionized hydrogen as the medium,
which is composed of only free electrons and protons. We introduce
the following $\kappa$ parameter to quantify the deviation from the
photon-like scenario: 
\begin{equation}
g_{Z'}^{(p)}:g_{Z'}^{(e)}=\kappa-1:1\thinspace.\label{eq:-21}
\end{equation}
Taking $\kappa=0$ or $1$ would correspond to a photon-like or baryophobic
 $Z'$. Using Eq.~\eqref{eq:-21} and repeating the calculations
in Sec.~\ref{subsec:vanishing}, we find that the final result in
Eq.~\eqref{eq:-20} for $\psi=e$ is changed to
\begin{equation}
\text{diagrams (a) + (b)}\propto g_{Z'}^{(e)}\frac{m_{Z'}^{2}-\kappa\Pi_{\gamma-(p)-\gamma}^{L,T}}{m_{Z'}^{2}-\Pi_{\gamma\gamma}^{L,T}}\thinspace,\label{eq:-22}
\end{equation}
which becomes insensitive to $m_{Z'}^{2}$ if $m_{Z'}^{2}$ is well
below $\kappa\Pi_{\gamma-(p)-\gamma}^{L,T}$ and $\Pi_{\gamma\gamma}^{L,T}$.
Substituting Eq.~\eqref{eq:-15} into Eq.~\eqref{eq:-22} and assuming
$\kappa\Pi_{\gamma-(p)-\gamma}^{L,T}<\Pi_{\gamma\gamma}^{L,T}$, we
obtain 
\begin{equation}
\text{diagrams (a) + (b)}\propto g_{Z'}^{(e)}\kappa\frac{m_{p}^{-1}}{m_{e}^{-1}+m_{p}^{-1}}\ \ \ \text{for}\ m_{Z'}^{2}\ll\kappa\frac{e^{2}n_{p}}{m_{p}}\thinspace.\label{eq:-23}
\end{equation}

In a homogeneous medium with infinite extent, the evolution of the
$Z'$ species is governed by the following Boltzmann equation:
\begin{equation}
\frac{df_{Z'}(\mathbf{k})}{dt}=\Gamma_{Z'}^{{\rm gain}}(\mathbf{k})\left(1+f_{Z'}(\mathbf{k})\right)-\Gamma_{Z'}^{{\rm loss}}(\mathbf{k})f_{Z'}(\mathbf{k})\thinspace,\label{eq:-24}
\end{equation}
where $f_{Z'}$ is the momentum distribution function of $Z'$, and
$\Gamma_{Z'}^{{\rm gain/loss}}$ is the gain/loss rate of $Z'$, to
be determined by evaluating collision terms for specific processes.
The $1+f_{Z'}$ factor attached to $\Gamma_{Z'}^{{\rm gain}}$ comes
from quantum statistics. 

For the photon $\gamma$, we define a similar gain/loss rate $\Gamma_{\gamma}^{{\rm gain/loss}}$
and $\Gamma_{\gamma}\equiv\Gamma_{\gamma}^{{\rm loss}}-\Gamma_{\gamma}^{{\rm gain}}$.
In the photon Boltzmann equation {[}similar to Eq.~\eqref{eq:-24}{]},
we have $df_{\gamma}/dt=0$ due to thermal equilibrium and hence
\begin{equation}
\Gamma_{\gamma}^{{\rm gain}}=f_{\gamma}\Gamma_{\gamma}\thinspace,\ \Gamma_{\gamma}^{{\rm loss}}=\left(1+f_{\gamma}\right)\Gamma_{\gamma}\thinspace.\label{eq:-37}
\end{equation}

Due to the weak couplings and low production rate, $Z'$ should be
far from reaching thermal equilibrium, i.e.~$f_{Z'}\ll1$. Hence
one can neglect the last term in Eq.~\eqref{eq:-24} and take $\Gamma_{Z'}^{{\rm gain}}\left(1+f_{Z'}\right)\approx\Gamma_{Z'}^{{\rm gain}}$
as the production rate. Under this assumption, Eq.~\eqref{eq:-24}
implies
\begin{equation}
\frac{dn_{Z'}}{dt}=\int\Gamma_{Z'}^{{\rm gain}}\frac{d^{3}\mathbf{k}}{(2\pi)^{3}}\thinspace,\label{eq:-27}
\end{equation}
where $n_{Z'}$ is the number density of $Z'$.

According to Eq.~\eqref{eq:-22}, $\Gamma_{Z'}^{{\rm gain}}$ can
be related to the gain rate of the photon $\Gamma_{\gamma}^{{\rm gain}}$
as follows:
\begin{equation}
\Gamma_{Z'}^{{\rm gain}}=\left|\frac{g_{Z'}^{(e)}}{eQ_{{\rm em}}^{(e)}}\frac{m_{Z'}^{2}-\kappa\Pi_{\gamma-(p)-\gamma}^{L,T}}{m_{Z'}^{2}-\Pi_{\gamma\gamma}^{L,T}}\right|^{2}\Gamma_{\gamma}^{{\rm gain}}\thinspace.\label{eq:-25}
\end{equation}
The photon gain rate $\Gamma_{\gamma}^{{\rm gain}}$ consists of two
dominant contributions, one from bremsstrahlung ($e^{-}+p\to e^{-}+p+\gamma$)
and the other from Thomson/Compton scattering ($\gamma+e^{-}\to\gamma+e^{-}$).
Including the two contributions, the explicit form of $\Gamma_{\gamma}^{{\rm gain}}$
in the longitudinal mode reads~\cite{Redondo:2013lna,Hardy:2016kme}:\footnote{Comparing to Eq.~(4.5) in Ref.~\cite{Redondo:2013lna}, we have
added an $f_{\gamma}$ factor because $\Gamma_{\gamma}^{{\rm gain}}=f_{\gamma}\Gamma_{\gamma}$.
Ref.~\cite{Hardy:2016kme} adopted the thermally-averaged Gaunt factor~\cite{Brussaard:1962zz}
to compute the bremsstrahlung contribution---see Eqs.~(A.5) and
(A.6) therein. We have checked that this is equivalent to Eq.~(4.5)
in Ref.~\cite{Redondo:2013lna} if the screening effect is negligible.
}

\begin{equation}
\Gamma_{\gamma}^{{\rm gain}}=f_{\gamma}\frac{64\pi^{2}\alpha^{3}n_{e}n_{p}}{3\sqrt{2\pi T}m_{e}^{3/2}\omega^{3}}F\left(\omega/T\right)+f_{\gamma}\frac{8\pi\alpha^{2}n_{e}}{3m_{e}^{2}}\sqrt{1-\frac{e^{2}n_{e}}{m_{e}\omega^{2}}}\Theta(\omega^{2}-e^{2}n_{e}/m_{e})\thinspace,\label{eq:-26}
\end{equation}
where $\alpha\equiv e^{2}/(4\pi)\approx1/137$, $f_{\gamma}=1/(e^{\omega/T}-1)$,
 $\Theta$ is the Heaviside theta function, and $F(x)\approx K_{0}(x/2)\sinh(x/2)$
assuming that the screening effect is negligible.   For the transverse
mode, the same expression in Eq.~\eqref{eq:-26} can be used at the
leading order~\cite{Redondo:2013lna}. 

The first and second terms in Eq.~\eqref{eq:-26} are proportional
to $\alpha^{3}$ and $\alpha^{2}$, corresponding to the contributions
of bremsstrahlung and Thomson/Compton scattering, respectively. It
is noteworthy that despite its higher order in $\alpha$, the first
term is generally greater than the second, because the initial states
of $e^{-}+p\to e^{-}+p+\gamma$ and $\gamma+e^{-}\to\gamma+e^{-}$
contain a proton and a photon, respectively, while the number density
of the former is much higher than the latter.  Taking the solar central
temperature $T\sim10^{7}K$ and density $\rho\sim150\text{g}/\text{cm}^{3}$~\cite{Xu:2022wcq}
for example, we obtain $n_{p}\approx\rho/m_{p}\approx9\times10^{25}/\text{cm}^{3}$
and $n_{\gamma}=2\zeta(3)T^{3}/\pi^{2}\approx2\times10^{22}/\text{cm}^{3}$,
i.e., $n_{p}$ is $\sim10^{3}$ higher than $n_{\gamma}$. This is
enough to compensate the difference between $\alpha^{3}$ and $\alpha^{2}$.

When using Eq.~\eqref{eq:-25} to compute $\Gamma_{Z'}^{{\rm gain}}$,
one may encounter resonance production occurring at $m_{Z'}^{2}\sim\Pi_{\gamma\gamma}^{L,T}$.
To the order of $\alpha$, $\Pi_{\gamma\gamma}^{L,T}$ are real and
given by Eq.~\eqref{eq:-15}. To the order of $\alpha^{2}$ or higher,
$\Pi_{\gamma\gamma}^{L,T}$ contain nonzero imaginary parts, which
can be determined using Weldon's formula~\cite{Weldon:1983jn}: 
\begin{equation}
{\rm Im}\Pi_{\gamma\gamma}^{L,T}=-\omega\Gamma_{\gamma}\thinspace,\label{eq:-44}
\end{equation}
where $\Gamma_{\gamma}$ can be computed using $\Gamma_{\gamma}^{{\rm gain}}=f_{\gamma}\Gamma_{\gamma}$
and Eq.~\eqref{eq:-26}. With Eq.~\eqref{eq:-44}, Eq.~\eqref{eq:-25}
can be rewritten as
\begin{equation}
\Gamma_{Z'}^{{\rm gain}}=C\frac{1}{\Delta^{2}+\Gamma_{\gamma}^{2}}\Gamma_{\gamma}\thinspace,\label{eq:-45}
\end{equation}
where $\Delta\equiv(m_{Z'}^{2}-{\rm Re}\Pi_{\gamma\gamma}^{L,T})/\omega$
and $C(\omega)$ absorbs unimportant quantities. The resonance occurs
at $\Delta=0$ and $\Gamma_{Z'}^{{\rm gain}}=C/\Gamma_{\gamma}$.
Despite that the height of the resonance is proportional to $1/\Gamma_{\gamma}$,
the overall contribution of the resonance to the integrated production
rate $\int\Gamma_{\gamma}^{{\rm gain}}d^{3}\mathbf{k}$ is insensitive
to $\Gamma_{\gamma}$, as previously pointed out in Ref.~\cite{An:2013yfc}.
This can be understood by noticing that at a wide range (e.g.~$\Delta\in[-K,\ K]$
with $K\gg\Gamma_{\gamma}$), Eq.~\eqref{eq:-45} behaves like the
Dirac delta function\footnote{Recall that the Dirac delta function can also be defined as $\delta(x)=\frac{1}{\pi}\lim_{\Gamma\to0}\frac{\Gamma}{x^{2}+\Gamma^{2}}$,
which is similar to the form of  Eq.~\eqref{eq:-45}. }, which is why the integrated contribution is insensitive to the height
of the resonance. \\

\begin{figure}
\centering

\includegraphics[width=0.95\textwidth]{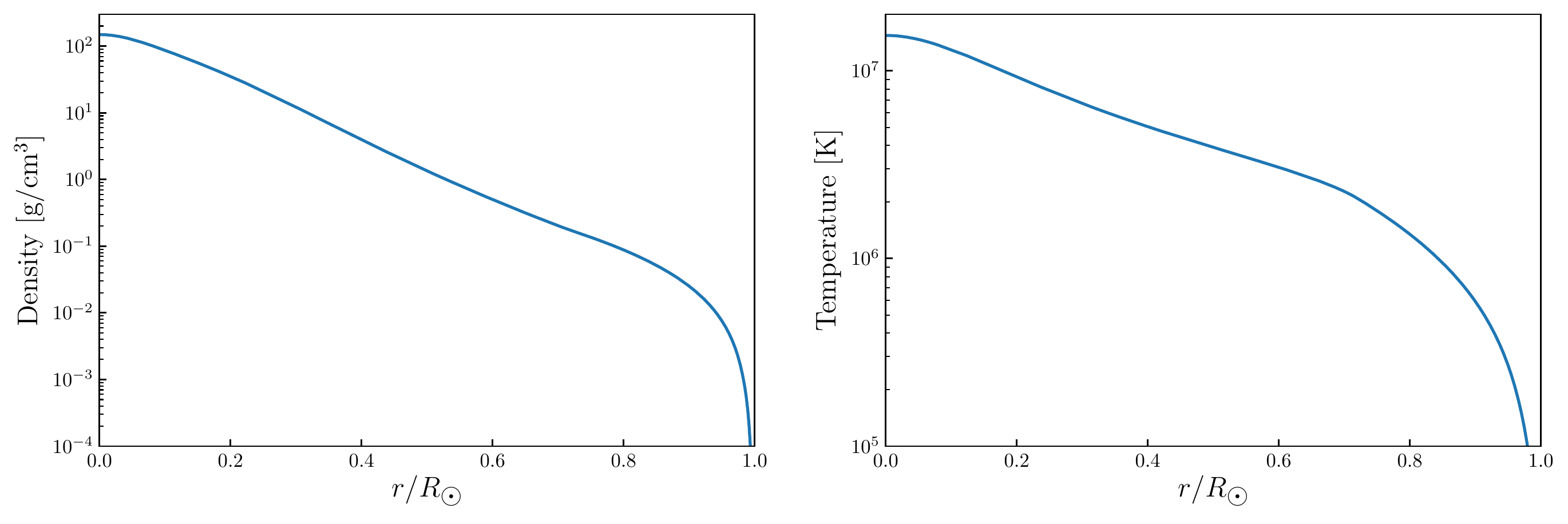}

\caption{\label{fig:solar-profile}The solar density and temperature profiles
used in this work. Data taken from Ref.~\cite{Vinyoles:2016djt}.}
\end{figure}

In the Sun, $Z'$ produced will immediately escape the finite extent
of the medium. In this case, Eq.~\eqref{eq:-27} should be interpreted
as the number of $Z'$ produced per unit volume per unit time in a
local region. The temperature and the density of the medium decrease
as the distance to the center $r$ increases. Fig.~\ref{fig:solar-profile}
shows the variation of the temperature and the density with respect
to $r$ in the standard solar model. The data is taken taken from
the latest calculation in Ref.~\cite{Vinyoles:2016djt} based on
the AGSS09 solar model~\cite{Asplund:2009fu}. 

The total number of $Z'$ produced in the Sun per unit time, $dN/dt$,
and the total energy loss per unit time, $dE_{{\rm loss}}/dt$, can
be obtained by integrating Eq.~\eqref{eq:-27} over the entire solar
profile:  
\begin{align}
\frac{dN}{dt} & =\int_{0}^{R_{\astrosun}}dr4\pi r^{2}\int_{0}^{\infty}dk\frac{k^{2}}{2\pi^{2}}\Gamma_{Z'}^{{\rm gain}}\thinspace,\label{eq:-28}\\
\frac{dE_{{\rm loss}}}{dt} & =\int_{0}^{R_{\astrosun}}dr4\pi r^{2}\int_{0}^{\infty}dk\frac{\omega k^{2}}{2\pi^{2}}\Gamma_{Z'}^{{\rm gain}}\thinspace.\label{eq:-29}
\end{align}

\begin{figure}
\centering

\includegraphics[width=0.8\textwidth]{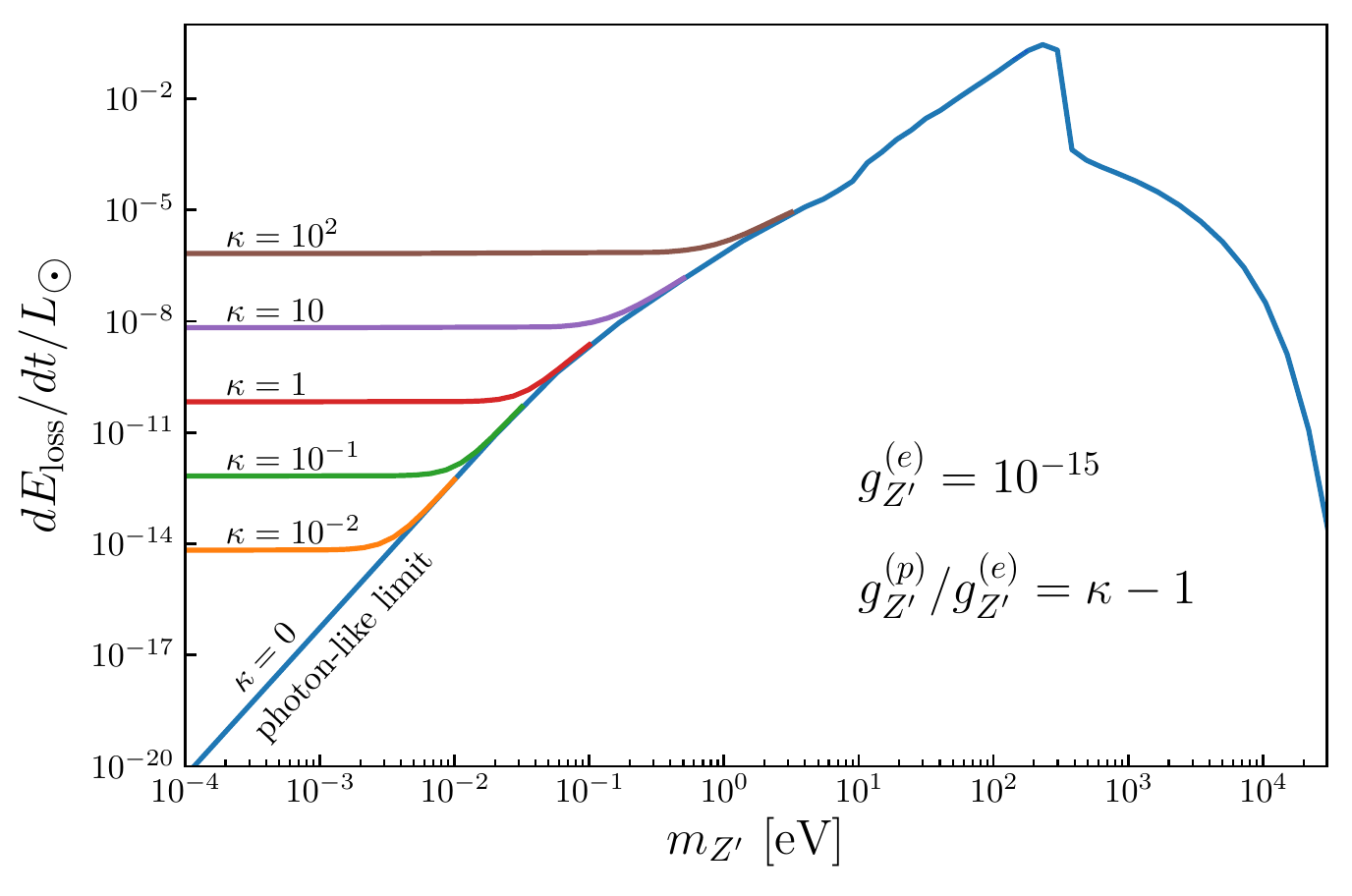}

\caption{\label{fig:prod_rate} The energy loss rate caused by $Z'$ emission
in the Sun for different values of $\kappa$, which is defined by
$g_{Z'}^{(p)}/g_{Z'}^{(e)}=\kappa-1$. The rate vanishes in the photon-like
limit ($\kappa=0$, $m_{Z'}\to0$) while increasing $\kappa$ leads
to a nonvanishing rate in the low-mass limit. }
\end{figure}

In Fig.~\ref{fig:prod_rate}, we show the energy loss rate $dE_{{\rm loss}}/dt$
computed using Eq.~\eqref{eq:-29} for a few selected values of $\kappa$
and $g_{Z'}^{(e)}=10^{-15}$. Eq.~\eqref{eq:-29} can be used to compute
the contribution of each polarization mode. In Fig.~\ref{fig:prod_rate}
we sum over one longitudinal and two transverse modes. The results
are normalized by the solar luminosity
\begin{equation}
L_{\astrosun}=3.83\times10^{26}\ \text{Watt}\thinspace.\label{eq:-38}
\end{equation}
As is expected from Eq.~\eqref{eq:-20}, the energy loss rate vanishes
in the photon-like limit ($\kappa=0$, $m_{Z'}\to0$) but if $\kappa\neq0$,
the curves in Fig.~\ref{fig:prod_rate} become flat below certain
mass scales, implying nonvanishing rates in the low-mass limit. 

\subsection{Including the neutron-$Z'$ coupling }

There is a considerably high abundance of neutrons in the Sun, mostly
in the form of $^{4}\text{He}$, $^{16}\text{O}$, $^{14}\text{C}$,
and other heavy nuclei. Overall, the ratio of the number densities
of neutrons and protons is $n_{n}/n_{p}\approx17\%$. Hence if $Z'$
is coupled to the neutron, its contribution is not negligible. To
include this contribution, we modify Eq.~\eqref{eq:-21} as follows
\begin{equation}
g_{Z'}^{(n)}:g_{Z'}^{(p)}:g_{Z'}^{(e)}=\kappa_{n}:\kappa-1:1\thinspace,\label{eq:-21-1}
\end{equation}
which introduces a new ratio $\kappa_{n}$. 

Note that  neutrons in the solar plasma are bound in nuclei and should
not be treated as free particles (the same for $\sim17\%$ of protons).
For simplicity, we assume all neutrons are bound in $^{4}\text{He}$
and neglect elements heavier than $^{4}\text{He}$. Under this assumption,
the plasma is composed of $e^{-}$, $p$, and ionized $^{4}\text{He}$.
Then Eq.~\eqref{eq:-22} is changed to 
\begin{equation}
\text{diagrams (a) + (b)}\propto g_{Z'}^{(e)}\frac{m_{Z'}^{2}-\kappa\Pi_{\gamma-(p)-\gamma}^{L,T}-\left(\kappa+\kappa_{n}\right)\Pi_{\gamma-({}^{4}\text{He})-\gamma}^{L,T}}{m_{Z'}^{2}-\Pi_{\gamma\gamma}^{L,T}}\thinspace,\label{eq:-22-1}
\end{equation}
where $\Pi_{\gamma-({}^{4}\text{He})-\gamma}^{L,T}$ takes the form
in Eq.~\eqref{eq:-15} with $n_{f}=n_{n}/2$ and $m_{f}=2m_{n}+2m_{p}\approx4m_{p}$. 

In particular, for the $B-L$ model ($\kappa_{n}=-1$, $\kappa=0$),
Eq.~\eqref{eq:-22-1} reduces to
\begin{equation}
\text{diagrams (a) + (b)}\propto g_{Z'}^{(e)}\frac{m_{Z'}^{2}+\Pi_{\gamma-({}^{4}\text{He})-\gamma}^{L,T}}{m_{Z'}^{2}-\Pi_{\gamma\gamma}^{L,T}}\thinspace,\label{eq:-39}
\end{equation}
which implies that the production rate is nonvanishing at $m_{Z'}\to0$
due to the presence of $^{4}\text{He}$ in the Sun. This feature has
been previously shown in Ref.~\cite{Hardy:2016kme}---see Fig.~4
therein. 

To compute the production rate of a generic $Z'$, we only need to
take Eq.~\eqref{eq:-25} with the numerator $m_{Z'}^{2}-\kappa\Pi_{\gamma-(p)-\gamma}^{L,T}$
replaced by the one in Eq.~\eqref{eq:-22-1}, and modify Eq.~\eqref{eq:-26}
with $n_{e}n_{p}\to n_{e}\sum_{f}Q_{f}^{2}n_{f}$ where $f$ in the
sum runs over all relevant nucleus species.

\section{Stellar cooling bounds for various $U(1)_{X}$ models\label{sec:bounds}}

The standard solar model has been well established with a variety
of predictions (e.g.~solar neutrino fluxes) in good agreement with
observations~\cite{Gann:2021ndb,Xu:2022wcq}. In the presence of
additional energy loss due to the emission of dark particles, many
of its  predictions could be altered. Without running dedicated simulations
for the altered solar models, it is generally believed that one can
take $dE_{{\rm loss}}/dt\lesssim10\%L_{\astrosun}$ as a conservative
bound on the energy loss rate~\cite{Gondolo:2008dd,Redondo:2013lna,Hardy:2016kme}.
In our work, we also take $10\%L_{\astrosun}$ as the maximally allowed
value of $dE_{{\rm loss}}/dt$ in the Sun. 

\begin{figure}
\centering

\includegraphics[width=0.98\textwidth]{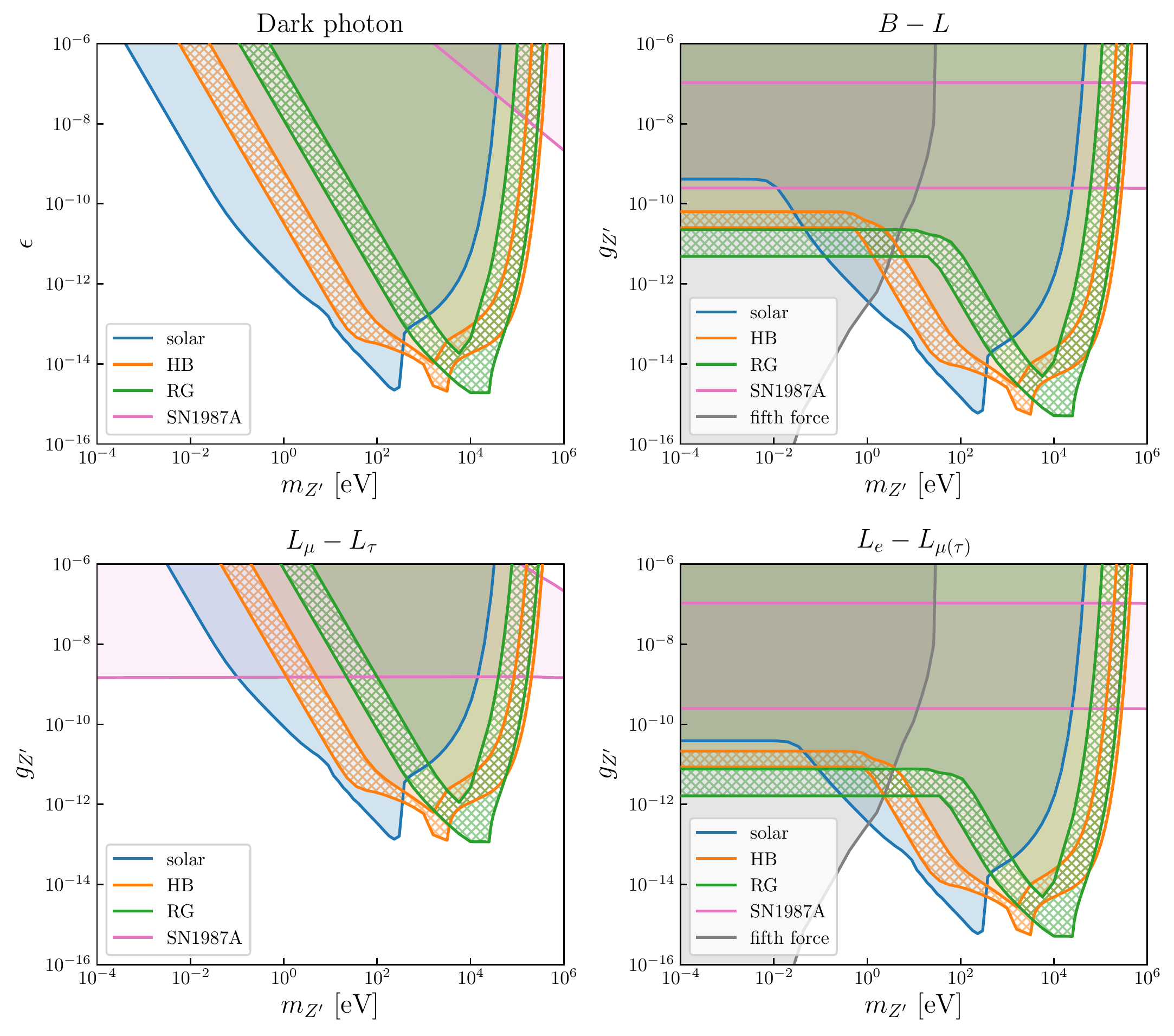}

\caption{\label{fig:constraint} Stellar (Solar, HB, RG) cooling bounds on
four $U(1)_{X}$ models compared with existing bounds (SN1987A and
fifth force). For the $L_{\mu}-L_{\tau}$ model, we have used loop-induced
couplings to electrons and protons---see Eq.~\eqref{eq:loop}. Curves
in the right panels become flat below certain mass scales because
$g_{Z'}^{(p)}/g_{Z'}^{(e)}\protect\neq-1$ and/or $g_{Z'}^{(n)}/g_{Z'}^{(e)}\protect\neq0$,
as discussed in the text. The supernova bounds (SN1987A) are taken
from Refs.~\cite{Chang:2016ntp,Croon:2020lrf}. The fifth-force bounds
consist of laboratory limits from the inverse-square-law test of gravity
and the Casimir effect, taken from Ref.~\cite{Heeck:2014zfa}. They
are absent in the left panels for photon-like couplings. }
\end{figure}

We consider a few popular $U(1)_{X}$ models including the dark photon
($Q_{X}=0$), the $B-L$ model which is the simplest flavor-universal
anomaly-free $U(1)$ extension of the SM, and leptonic $U(1)$ extensions
$L_{\alpha}-L_{\beta}$ with $\alpha,\ \beta\in\{e,\ \mu,\ \tau\}$.
The $L_{\mu}-L_{\tau}$ model has gained rising interest due to its
viability to accommodate the muon $g-2$ anomaly after combining all
known bounds. Although $Z'$ in this model does not couple to electrons
and quarks directly, there are loop-induced  couplings to all charged
fermions~\cite{Araki:2017wyg}:
\begin{equation}
g_{Z'}^{(f)}=-\frac{\alpha}{3\pi}g_{Z'}Q_{{\rm em}}^{(f)}\log\left(\frac{m_{\mu}^{2}}{m_{\tau}^{2}}\right)\thinspace.\label{eq:loop}
\end{equation}
Note that the loop-induced couplings are proportional to $Q_{{\rm em}}^{(f)}$,
which implies that these couplings are photon-like and hence the production
rate vanishes at $m_{Z'}\to0$. Eq.~\eqref{eq:loop} is generated
by loop diagrams with an internal photon mediator and a $\mu/\tau$
fermion loop. If the photon mediator is replaced by the SM $Z$ boson,
one would get more suppressed loop-induced couplings. They are suppressed
by $m_{Z'}^{2}/m_{Z}^{2}$ and hence negligible in our analysis. 

For these models, we take the following values for $\kappa$ and $\kappa_{n}$:
\begin{align}
\text{dark photon} & :\ (\kappa,\ \kappa_{n})=(0,\ \ \ 0),\ \ \ \ \ g_{Z'}^{(e)}=ec_{W}\epsilon\thinspace,\label{eq:-40}\\
B-L & :\ (\kappa,\ \kappa_{n})=(0,\ -1),\ \ \thinspace\ \ g_{Z'}^{(e)}=-g_{Z'}\thinspace,\label{eq:-41}\\
L_{\mu}-L_{\tau} & :\ (\kappa,\ \kappa_{n})=(0,\ \ \ 0),\ \ \ \ \ g_{Z'}^{(e)}=-4.4\cdot10^{-3}g_{Z'}\thinspace,\label{eq:-42}\\
L_{e}-L_{\mu(\tau)} & :\ (\kappa,\ \kappa_{n})=(1,\ \ \ 0),\ \ \ \ \ g_{Z'}^{(e)}=g_{Z'}\thinspace.\label{eq:-43}
\end{align}
In Eq.~\eqref{eq:-43}, we assume that the loop-induced couplings
are negligible (if included, they change $\kappa$ to $0.992$ and
$0.987$ for $L_{e}-L_{\mu}$ and $L_{e}-L_{\tau}$, respectively)
so that $L_{e}-L_{\mu}$ and $L_{e}-L_{\tau}$ are treated as the
same model in our analysis. In Eq.~\eqref{eq:-40}, $\kappa$ and
$\kappa_{n}$ in principle should slightly deviate from $0$ according
to Tab.~\ref{tab:g-eff}, with the deviations proportional to $r_{m}=m_{Z'}^{2}/m_{Z}^{2}$.
However, since $r_{m}$ is extremely small ($\sim10^{-22}$ for $m_{Z'}\sim1$
eV), including the small deviation would not cause visible changes
in our final results.  

By requiring that the energy loss rate is below the maximally allowed
limit, we obtain the constraints on $\epsilon$ or $g_{Z'}$ for these
models in Fig.~\ref{fig:constraint}. As is expected from our previous
discussions, when $m_{Z'}$ approaches zero, the stellar cooling bounds
vanish for the dark photon model and the $L_{\mu}-L_{\tau}$ model
because the couplings in these models are photon-like. The bounds
for $B-L$ and $L_{e}-L_{\mu(\tau)}$, by contrast,  become flat below
certain mass scales.  This is caused by $\kappa_{n}\neq0$ in the
$B-L$ model and $\kappa\neq0$ in the $L_{e}-L_{\mu(\tau)}$ model.
Our results indicate that stellar cooling constraints on generic $Z'$
in the low-mass limit crucially depend on whether the couplings to
the medium particles are in alignment with those of the photon. 

In addition to the Sun, one can readily apply the above calculations
to other stars with much higher core temperatures and core densities
such as red giants (RG) and horizontal branch (HB) stars~\cite{Raffelt1996}\footnote{See Chapter 2.1.3, Figs.~2.4 and 2.6 therein.}.
For RG, we consider the tip of their evolution before He ignition.
At this tip, the orders of magnitude of the core temperature and density
are $T\sim10^{8}\ \text{K}$ and $\rho\sim10^{6}\ \text{g}/\text{cm}^{3}$,
while the specific values depend on model-specific simulations. According
to several samples of such simulations~\cite{Passy:2012uf,Dessert:2021wjx},
we vary the core temperature and density in the range $T\in[10^{7.5},\ 10^{8.0}]\ \text{K}$
and $\rho\in[10^{5.5},\ 10^{6.0}]\ \text{g}/\text{cm}^{3}$. This
corresponds to the green bands in Fig.~\ref{fig:constraint}. For
HB stars in which the ignited He is burning, the core density decreases
to $\rho\sim10^{4}\text{g}/\text{cm}^{3}$ while the core temperature
remains roughly the same after He ignition. According to Ref~\cite{Raffelt1996},
we vary the core temperature and density in the range $T\in[0.6,\ 1.2]\times10^{8}\ \text{K}$
and $\rho\in[10^{3.6},\ 10^{4.2}]\ \text{g}/\text{cm}^{3}$, corresponding
to the orange bands in Fig.~\ref{fig:constraint}. For both RG and
HB stars, we assume that the energy loss rate at the core should be
less than $10\ \text{erg}\ \text{g}^{-1}\text{sec}^{-1}$ to maintain
the agreement between the standard theory and the observations. The
chemical composition at the core is assumed to be dominated by He.

In Fig.~\ref{fig:constraint}, we also include existing bounds from
the fifth-force searches~\cite{Heeck:2014zfa} and the observation
of supernova 1987A~\cite{Chang:2016ntp,Croon:2020lrf} for comparison.
The supernova bound for the dark photon model, taken from~\cite{Chang:2016ntp},
exhibits a very different low-mass limit compared to those taken from~\cite{Croon:2020lrf}.
This is due to the photon-like coupling and the cancellation previously
discussed. For $L_{\mu}-L_{\tau}$, this cancellation could also be
present in supernovae but it should disappear when supernova muons
are taken into account~\cite{Croon:2020lrf}. As for bounds from
the fifth-force searches, they cannot be applied to photon-like couplings
due to the cancellation between positive and negative charge contributions
in normal matter. 

In the presented mass range, there could also be other laboratory
bounds from e.g. neutrino-electron scattering~\cite{Bilmis:2015lja,Lindner:2018kjo}
and electron and muon $g-2$ (see e.g.~\cite{Bauer:2018onh}). These
bounds are only relevant when $g_{Z'}$ or $\epsilon$ is at least
above $10^{-5}$ so they are not included in the figure. Beam dump
bounds~(see e.g.~\cite{Coy:2021wfs}) are irrelevant when $m_{Z'}<2m_{e}$
because $Z'$ cannot decay to the lightest charged particle.

\section{Conclusions\label{sec:Con}}

Dark photons and other dark gauge bosons, generically denoted by
$Z'$ in this work, could be produced in a hot and dense stellar medium,
causing additional energy loss which has been used to derive one of
the most constraining bounds on these hypothetical particles. 

We show that the production rate of $Z'$ in the low-mass limit crucially
depends on how close its couplings are to the photon-like limit, as
illustrated in Fig.~\ref{fig:prod_rate}. Consequently, the stellar
cooling bounds on $Z'$ from different models can be very different,
as shown in Fig.~\ref{fig:constraint}. 

For $B-L$ and $L_{e}-L_{\mu(\tau)}$ models with $m_{Z'}\lesssim10\ \text{keV}$,
the gauge coupling $g_{Z'}$ is constrained by the solar model to
be at least below $4.1\times10^{-10}$ and $3.9\times10^{-11}$, respectively.
The constraints would be much stronger if $m_{Z'}$ is in the resonant
production region,  or if one considers RG and HB stars, albeit subject
to more astrophysical uncertainties. For dark photon and $L_{\mu}-L_{\tau}$
models, the effective couplings generated via kinetic mixing or loop
diagrams are photon-like. Hence the lower bounds on $\epsilon$ or
$g_{Z'}$ for these two models are always mass-dependent. 

 Our model-specific bounds derived from stellar cooling  might be
of importance to  a variety of studies on new light particles at
the intensity frontier and in cosmology. 
\begin{acknowledgments}
We would like to thank Xuheng Luo for inspiring discussions on dark
photon mixing in a medium, and Evgeny Akhmedov for helpful clarification
on photon scattering issues. This work is supported in part by the
National Natural Science Foundation of China under grant No. 12141501. 
\end{acknowledgments}

\appendix

\section{Transformation from the original basis to the physical basis\label{sec:Transformation}}

In this appendix, we present the detail of how the original basis
is transformed to the physical basis. As has been briefly described
below Eq.~\eqref{eq:L}, the transformation involves two steps. First
one needs to canonicalize the kinetic terms via a linear non-unitary
transformation, denoted by $\mathbb{L}$, and then diagonalize the
mass matrix of gauge bosons by a unitary transformation, denoted by
$\mathbb{O}$. The two steps are formulated as follows: 
\begin{equation}
\hat{\mathbf{V}}\xrightarrow{\mathbb{L}}\mathbf{V}\xrightarrow{\mathbb{O}}\mathbf{V}_{{\rm ph}}\thinspace,\ \ \mathbf{V}=\mathbb{L}\hat{\mathbf{V}},\ \mathbf{V}_{{\rm ph}}=\mathbb{O}\mathbf{V}\thinspace,\label{eq:-13}
\end{equation}
where 
\begin{align}
\hat{\mathbf{V}} & \equiv(W_{1},\ W_{2},\ W_{3},\ \hat{B},\ \hat{X})^{T}\thinspace,\label{eq:-1}\\
\mathbf{V} & \equiv(W_{1},\ W_{2},\ W_{3},\ B,\ X)^{T}\thinspace,\label{eq:-2}\\
\mathbf{V}_{{\rm ph}} & \equiv(W_{1},\ W_{2},\ A,\ Z,\ Z')^{T}\thinspace.\label{eq:-3}
\end{align}
    The specific forms of $\mathbb{L}$ and $\mathbb{O}$ in Eq.~\eqref{eq:-13}
are given as follows~\cite{Lindner:2018kjo}:
\begin{equation}
\mathbb{L}=\left(\begin{array}{ccc}
\mathbb{I}_{3}\\
 & 1 & \epsilon\\
 & 0 & \sqrt{1-\epsilon^{2}}
\end{array}\right),\ \mathbb{O}=\left(\begin{array}{cccc}
\mathbb{I}_{2}\\
 & s_{W} & c_{W} & 0\\
 & c_{W}c_{\theta} & -s_{W}c_{\theta} & -s_{\theta}\\
 & c_{W}s_{\theta} & -s_{W}s_{\theta} & c_{\theta}
\end{array}\right),\label{eq:-4}
\end{equation}
where $\mathbb{I}_{n}$ is an $n\times n$ identity matrix, $(s_{W}\ c_{W})\equiv(\sin\theta_{W},\ \cos\theta_{W})$
with $\theta_{W}$ the Weinberg angle, and $(s_{\theta}\ c_{\theta})\equiv(\sin\theta,\ \cos\theta)$.
The angle $\theta$, to be determined later, is similar to $\theta_{W}$
in the sense that it describes how much the neutral massive boson
of $SU(2)_{L}$ mixes with $U(1)_{X}$ instead of $U(1)_{Y}$. Hence
the orthogonal transformation $\mathbb{O}$ can be regarded as a generalized
Weinberg rotation in the $SU(2)_{L}\times U(1)_{Y}\times U(1)_{X}$
model, with two angles instead of one. 

The $\mathbb{T}$ matrix in Eq.~\eqref{eq:-30} is given by 
\begin{equation}
\mathbb{T}=\mathbb{O}\cdot\mathbb{L}\thinspace.\label{eq:-46}
\end{equation}
Note that the specific forms of $\mathbb{L}$ and $\mathbb{O}$ are
not unique because  $(\mathbb{L},\ \mathbb{O})\to(\mathbb{L}',\ \mathbb{O}')=(\mathbb{A}\cdot\mathbb{L},\ \mathbb{O}\cdot\mathbb{A}^{T})$
with an arbitrary orthogonal matrix $\mathbb{A}$ ($\mathbb{A}\cdot\mathbb{A}^{T}=1$)
can always canonicalize the kinetic terms and diagonalize the mass
matrix. Only their product $\mathbb{T}=\mathbb{O}\cdot\mathbb{L}=\mathbb{O}'\cdot\mathbb{L}'$
is unique. 

The angle $\theta$  depends on the mass matrix of gauge bosons. If
the $U(1)_{X}$ gauge boson acquires a mass via the Stückelberg mechanism
or some dark-sector Higgses which are SM singlets, then there is no
mass mixing between $\hat{X}$ and $\hat{B}$, but the $\mathbb{L}$
transformation causes mass mixing between $X$ and $B$. After the
$\mathbb{O}$ transformation, the mass matrix is re-diagonalized and
$\theta$ is determined by~\cite{Lindner:2018kjo}
\begin{equation}
\tan\theta=\frac{\epsilon m_{Z}^{2}s_{W}}{m_{Z'}^{2}-m_{Z}^{2}}+{\cal O}(\epsilon^{2})\thinspace.\label{eq:-5}
\end{equation}

If the $U(1)_{X}$ gauge boson acquires its mass via Higgs doublets,
 then there would be mass mixing in the $\hat{\mathbf{V}}$ basis.
After the $\mathbb{L}$ and $\mathbb{O}$ transformations, $\tan\theta$
receives an additional contribution almost independent of $\epsilon$,
i.e. the mass mixing depends not only on $\epsilon$ but also on additional
parameters.  In this case, we can treat $\theta$ as a free parameter.

The gauge interactions in the $\hat{\mathbf{V}}$ basis read
\begin{equation}
{\cal L}\supset\sum_{\psi}\overline{\psi}\gamma^{\mu}\mathbf{Q}\hat{\mathbf{V}}_{\mu}\psi\thinspace,\label{eq:-6}
\end{equation}
where 
\begin{equation}
\mathbf{Q}\equiv\left(g\frac{\sigma_{1}}{2},\ g\frac{\sigma_{2}}{2},\ g\frac{\sigma_{3}}{2},\ g'Q_{Y},\ g_{Z'}Q_{X}\right).\label{eq:-7}
\end{equation}
After the transformation $\hat{\mathbf{V}}\to\mathbf{V}_{{\rm ph}}=\mathbb{T}\hat{\mathbf{V}}$,
one obtains the gauge interactions in the physical ($\mathbf{V}_{{\rm ph}}$)
basis. It is straightforward to see that the charged-current interactions
in the SM are not modified. So we are only concerned with the interactions
of $A$, $Z$, and $Z'$. The result is
\begin{equation}
{\cal L}\supset eQ_{{\rm em}}^{(\psi)}\overline{\psi}\slashed{A}\psi+g_{Z}^{(\psi)}\overline{\psi}\slashed{Z}\psi+g_{Z'}^{(\psi)}\overline{\psi}\slashed{Z}'\psi\thinspace,\label{eq:-8}
\end{equation}
 which implies that the electromagnetic interactions remain the same
as in the SM. The effective couplings $g_{Z}^{(\psi)}$ and $g_{Z'}^{(\psi)}$
are given by
\begin{align}
g_{Z}^{(\psi)} & =c_{\theta}g\left[c_{W}Q_{{\rm em}}^{(\psi)}-\frac{Q_{Y}^{(\psi)}}{c_{W}}\right]-s_{\theta}\left[g_{Z'}Q_{X}^{(\psi)}-gQ_{Y}^{(\psi)}\frac{s_{W}}{c_{W}}\epsilon\right]+{\cal O}(\epsilon^{2}),\label{eq:-9}\\
g_{Z'}^{(\psi)} & =s_{\theta}g\left[c_{W}Q_{{\rm em}}^{(\psi)}-\frac{Q_{Y}^{(\psi)}}{c_{W}}\right]+c_{\theta}\left[g_{Z'}Q_{X}^{(\psi)}-gQ_{Y}^{(\psi)}\frac{s_{W}}{c_{W}}\epsilon\right]+{\cal O}(\epsilon^{2}).\label{eq:-10}
\end{align}
If the kinetic mixing term is the only source of mass mixing, then
we can apply Eq.~\eqref{eq:-5} to obtain the expressions of $g_{Z}^{(\psi)}$
and $g_{Z'}^{(\psi)}$ in Tab.~\ref{tab:g-eff}. 

\section{Medium-induced \label{sec:coherent-scattering}kinetic mixing rederived
from coherent scattering}

As we have discussed in Sec.~\ref{subsec:vanishing}, coherent scattering
of $\gamma$/$Z'$ with charged particles in a medium causes medium-induced
mixing between $\gamma$ and $Z'$. In finite-temperature/density
field theories, this can be obtained by computing the following diagram\begin{equation}
\begin{tikzpicture}
\begin{scope}[thick,decoration={
    markings,
    mark=at position 0.2 with {\arrow{stealth}}}
    ]

%\draw (0.5,0) [postaction={decorate}]

\draw (0.5,0)  arc
	[
		start angle=0,
		end angle=360,
		x radius=0.5,
		y radius =0.5
	] ;

%\draw [postaction={decorate}]%(-1.5,1.5)-- (-0.5,0);
%\draw[postaction={decorate}] (-0.5,0)--(-1.5,-1.5);

\draw[decorate,decoration=snake,segment amplitude=.6mm, segment length=2mm]  (-0.5,0)--(-1.5,0);

\draw[decorate,decoration=snake,segment amplitude=.6mm, segment length=2mm]  (0.5,0)--(1.5,0);

\node [text=black,rotate=0] at (-1.4,0.3) {$\gamma$};
\node [text=black,rotate=0] at (1.4,0.3) {$Z'$};
\node [text=black,rotate=0] at (-0,0.25) {$f$}; 
\node [text=black,rotate=0] at (3.0,0) {$=\Pi^{\mu\nu}_{\gamma-(f)-Z'}\ ,$};

\end{scope}

\end{tikzpicture}
\end{equation}using medium-modified fermion propagators in the loop. And the result
is essentially the well-known plasmon masses given in Eq.~\eqref{eq:-15}
with $Q_{f}^{2}e^{2}\to Q_{f}eg_{Z'}^{(f)}$. 

Here we would like to rederive it from the coherent scattering of
$\gamma$/$Z'$ with medium particles. By directly computing the following
diagrams \begin{equation}
\begin{tikzpicture}
\begin{scope}[thick,decoration={
    markings,
    mark=at position 0.6 with {\arrow{stealth}}}
    ]

\draw[postaction={decorate}] (-1.5,-1.0)--(-0.5,0);
\draw[postaction={decorate}] (0.5,0)--(1.5,1.0);
\draw[postaction={decorate}] (-0.5,0)--(0.5,0);

\draw[decorate,decoration=snake,segment amplitude=.6mm, segment length=2mm]  (-0.5,0)--(-1.5,0);

\draw[decorate,decoration=snake,segment amplitude=.6mm, segment length=2mm]  (0.5,0)--(1.5,0);

\node [text=black,rotate=0] at (-1.4,0.3) {$\gamma$};
\node [text=black,rotate=0] at (1.4,-0.3) {$Z'$};
\node [text=black,rotate=0] at (-0.8,-0.6) {$f$}; 
\node [text=black,rotate=0] at (2.0,0) {$+$}; 

\end{scope}
\end{tikzpicture}
\begin{tikzpicture}
\begin{scope}[thick,decoration={
    markings,
    mark=at position 0.6 with {\arrow{stealth}}}
    ]

\draw[postaction={decorate}] (-0.5,-1.0)--(0.5,0);
\draw[postaction={decorate}] (-0.5,0)--(0.5,1.0);
\draw[postaction={decorate}] (0.5,0)--(-0.5,0);

\draw[decorate,decoration=snake,segment amplitude=.6mm, segment length=2mm]  (-0.5,0)--(-1.5,0);

\draw[decorate,decoration=snake,segment amplitude=.6mm, segment length=2mm]  (0.5,0)--(1.5,0);

\node [text=black,rotate=0] at (-1.4,0.3) {$\gamma$};
\node [text=black,rotate=0] at (1.4,-0.3) {$Z'$};
\node [text=black,rotate=0] at (-0.4,-0.6) {$f$}; 
\node [text=black,rotate=0] at (3.0,0) {$={\cal M}_{\mu\nu}\epsilon_{\gamma}^{\mu}\epsilon_{Z'}^{\nu}\ ,$};

\end{scope}

\end{tikzpicture}
\label{eq:feyn2}
\end{equation} and assuming that the scattering is coherent among multiple medium
particles in a local region, one can obtain the same result, but the
rederivation provides a more intuitive understanding of the medium
effect. 

Let us start with Eq.~\eqref{eq:feyn2} where $\epsilon_{\gamma}$
and $\epsilon_{Z'}$ denote the polarization vectors of $\gamma$
and $Z'$ respectively. The fermionic part of  Eq.~\eqref{eq:feyn2}
reads
\begin{equation}
{\cal M}^{\mu\nu}=eQ_{f}g_{Z'}^{(f)}\overline{u}\left[\gamma^{\nu}\frac{1}{\slashed{p}_{a}-m_{f}}\gamma^{\mu}+\gamma^{\mu}\frac{1}{\slashed{p}_{b}-m_{f}}\gamma^{\nu}\right]u\thinspace,\label{eq:-47}
\end{equation}
where $p_{a}$ and $p_{b}$ denote the momentum of the internal fermion
propagator of the first and second diagrams respectively. As a condition
of coherence, the momentum transfer from the photon to the medium
particle should be sufficiently small. It should be well below the
inverse of the radius of the local region maintaining coherency. For
simplicity, we take the zero limit of the momentum transfer so that
the initial $\gamma$ and the final $Z'$ have exactly the same momentum.
Under this limit, we have $p_{a}^{\mu}=p^{\mu}+k^{\mu}$ and $p_{b}^{\mu}=p^{\mu}-k^{\mu}$
where $k^{\mu}$ and $p^{\mu}$ are the momenta of $\gamma$ and $f$
respectively. Then after applying on-shell conditions ($p^{2}=m_{f}^{2}$,
$\slashed{p}u=m_{f}u$, $\overline{u}\slashed{p}=\overline{u}m_{f}$),
Eq.~\eqref{eq:-47} becomes
\begin{equation}
{\cal M}^{\mu\nu}=eQ_{f}g_{Z'}^{(f)}\frac{4k.p\left(j^{\nu}k^{\mu}+j^{\mu}k^{\nu}-j\cdot k\thinspace g^{\mu\nu}\right)-2k^{2}\left(j^{\nu}p^{\mu}+j^{\mu}p^{\nu}\right)}{4(k\cdot p)^{2}-(k^{2})^{2}}\thinspace,\label{eq:-48}
\end{equation}
with 
\begin{equation}
j^{\mu}\equiv\overline{u}\gamma^{\mu}u\thinspace.\label{eq:-49}
\end{equation}

Given the amplitude ${\cal M}^{\mu\nu}$, it is straightforward to
obtain the corresponding effective Lagrangian in the momentum space:
\begin{equation}
{\cal L}_{{\rm eff}}={\cal M}^{\mu\nu}|_{j\to J}A_{\mu}Z'_{\nu}\thinspace,\label{eq:-51}
\end{equation}
where
\begin{equation}
J_{\mu}\equiv\overline{f}\gamma_{\mu}f\thinspace.\label{eq:-52}
\end{equation}
Note that $J$ and $j$ have different dimensions, $J\sim[E]^{3}$
vs $j\sim[E]^{1}$.  In a background of a large number of $f$ particles,
the expectation value $\langle J_{\mu}\rangle$ is exactly the classical
current of $f$ particles. In particular, for a non-relativistic medium,
we have 
\begin{equation}
\langle J_{\mu}\rangle\approx(n_{f},\ 0,\ 0,\ 0)\thinspace,\label{eq:-53}
\end{equation}
where $n_{f}$ is the number density of $f$. Coherent scattering
of $\gamma$/$Z'$ with the background particles leads to a background
expectation value of ${\cal M}^{\mu\nu}|_{j\to J}$, which gives rise
to the medium-induced mixing between $\gamma$ and $Z'$:
\begin{equation}
\Pi_{\gamma-(f)-Z'}^{\mu\nu}=\left\langle {\cal M}^{\mu\nu}|_{j\to J}\right\rangle .\label{eq:-54}
\end{equation}

Next, we use Eqs.~\eqref{eq:-48} and \eqref{eq:-53} to compute $\Pi_{\gamma-(f)-Z'}^{\mu\nu}$
in Eq.~\eqref{eq:-54}. Since ${\cal M}^{\mu\nu}$ is multiplied by
$\epsilon_{\gamma}^{\mu}\epsilon_{Z'}^{\nu}$ and $k\cdot\epsilon_{\gamma}=k\cdot\epsilon_{Z'}=0$,
the $j^{\nu}k^{\mu}+j^{\mu}k^{\nu}$ term in Eq.~\eqref{eq:-48} can
be ignored. In addition, we only consider a non-relativistic medium,
which implies 
\begin{equation}
p^{\mu}\approx(m_{f},\ 0,\ 0,\ 0)\thinspace.\label{eq:-56}
\end{equation}
Since the photon is soft compared to the fermion mass ($k\ll m_{f}$),
we can neglect $(k^{2})^{2}$ in the denominator of Eq.~\eqref{eq:-48}.
From Eqs.~\eqref{eq:-56} and \eqref{eq:-53}, we have $\langle J\rangle\cdot k/\left(p\cdot k\right)=n_{f}/m_{f}$.
Combining all these together, we obtain 
\begin{equation}
\Pi_{\gamma-(f)-Z'}^{\mu\nu}=eQ_{f}g_{Z'}^{(f)}\left[-\frac{n_{f}}{m_{f}}g^{\mu\nu}-\frac{k^{2}}{2(k\cdot p)^{2}}\left(\langle J^{\nu}\rangle p^{\mu}+\langle J^{\mu}\rangle p^{\nu}\right)\right].\label{eq:-50}
\end{equation}

Without loss of generality, one can assume that $k^{\mu}$ is in the
$z$-axis direction, $k^{\mu}=(\omega,\ 0,\ 0,\ |\mathbf{k}|)$ so
that the polarization vectors in Eq.~\eqref{eq:-16} can be written
as follows:
\begin{equation}
\epsilon_{T1}^{\mu}=(0,\ 1,\ 0,\ 0)\thinspace,\ \epsilon_{T2}^{\mu}=(0,\ 0,\ 1,\ 0)\thinspace,\ \epsilon_{L}^{\mu}=\frac{1}{\sqrt{k^{2}}}(|\mathbf{k}|,\ 0,\ 0,\ \omega)\thinspace.\label{eq:-55}
\end{equation}
Each of them satisfies $k\cdot\epsilon=0$ and $\epsilon^{\mu}\epsilon_{\mu}=-1$. 

Applying the decomposition in Eq.~\eqref{eq:-16}, we obtain 
\begin{align}
\Pi_{\gamma-(f)-Z'}^{T} & =eQ_{f}g_{Z'}^{(f)}\frac{n_{f}}{m_{f}}\thinspace,\label{eq:-57}\\
\Pi_{\gamma-(f)-Z'}^{L} & =eQ_{f}g_{Z'}^{(f)}\frac{n_{f}}{m_{f}}\left[1-\frac{|\mathbf{k}|^{2}}{\omega^{2}}\right],\label{eq:-58}
\end{align}
where we have used $\epsilon_{T1}^{\mu}p_{\mu}=0$, $\epsilon_{T2}^{\mu}p_{\mu}=0$,
$\epsilon_{L}^{\mu}p_{\mu}=m_{f}|\mathbf{k}|/\sqrt{k^{2}}$, and $\epsilon_{L}^{\mu}\langle J_{\mu}\rangle=n_{f}|\mathbf{k}|/\sqrt{k^{2}}$.
 Therefore, the medium-induced kinetic mixing derived from the coherent
scattering theory is identical to that obtained using the finite-temperature/density
field theory. 

\bibliographystyle{JHEP}
\bibliography{ref}

\end{document}